\documentclass[aps,showpacs,superscriptaddress,notitlepage]{revtex4-1}
\usepackage{amsmath}
\usepackage{epsfig}
\def\bea#1\eea{\begin{align}#1\end{align}}
\newcommand{\nnu}{\nonumber\\}
\newcommand{\bef}{\begin{figure}[htb]\centering}
\newcommand{\eef}{\end{figure}}

\begin{document}
\title{Multiple scattering effects on inclusive particle production in the large-$x$ regime}

\date{\today}

\author{Zhong-Bo Kang}
\email{zkang@lanl.gov}
\affiliation{Theoretical Division, 
                   Los Alamos National Laboratory, 
                   Los Alamos, NM 87545, USA}
                   
\author{Ivan Vitev}
\email{ivitev@lanl.gov}
\affiliation{Theoretical Division, 
                   Los Alamos National Laboratory, 
                   Los Alamos, NM 87545, USA}                   

\author{Hongxi Xing}
\email{xinghx@iopp.ccnu.edu.cn}
\affiliation{Theoretical Division, 
                   Los Alamos National Laboratory, 
                   Los Alamos, NM 87545, USA}
\affiliation{Interdisciplinary Center for Theoretical Study and Department of Modern Physics,
                   University of Science and Technology of China, 
                   Hefei 230026, China}
\affiliation{Institute of Particle Physics, 
                   Central China Normal University, 
                   Wuhan 430079, China}

\begin{abstract}
We study the multiple scattering effects on inclusive  particle production in p+A and $\gamma$+A collisions. Specifically, 
we concentrate on the region where the parton momentum fraction in the nucleus $x\sim {\cal O}(1)$ and 
 incoherent multiple interactions are relevant. By taking into account both initial-state and final-state double 
scattering, we derive the nuclear size-enhanced power corrections to the differential cross section for single inclusive 
hadron production in p+A and $\gamma$+A reactions, and for prompt photon production in p+A reactions. We find that the 
final result can be written in a simple compact form in terms of four-parton correlation functions, in which 
the second-derivative, first-derivative and non-derivative terms of the correlation distributions share the same 
hard-scattering functions. We expect our result to be especially relevant for understanding the nuclear modification of
particle production in the backward rapidity regions  in p+A and e+A collisions. 
\end{abstract}

\maketitle

\section{Introduction}
Medium-induced modification of moderate and high transverse momentum particle production in both proton-nucleus (p+A) 
and nucleus-nucleus (A+A) collisions relative to the naive binary collision-scaled proton-proton (p+p) baseline 
expectation is an excellent tool to diagnose the properties of dense QCD 
matter~\cite{Gyulassy:2003mc,Albacete:2013ei}. Multiple  parton scattering has played an important role in understanding 
novel effects that contribute to the observed nuclear dependence, such as dynamical shadowing, Cronin effect, parton 
energy  loss and jet  quenching~\cite{Vitev:2006bi}. To better extract the QCD matter properties from experimental 
measurements,  it is critical to elucidate the differences between the elastic, inelastic, coherent, and incoherent 
scattering regimes.

Different  theoretical  approaches are possible in studying the same physics effect in  high energy nuclear collisions. 
An illustrative example is the calculation of medium-induced parton splitting and radiative energy loss 
that  leads to the  ``jet quenching'' phenomena  in A+A 
reactions~\cite{Baier:1996kr,Zakharov:1997uu,Gyulassy:2000fs,Wiedemann:2000za,Guo:2000nz,Arnold:2002ja,Ovanesyan:2011xy}. 
In p+A collisions, most attention has been devoted  to the non-trivial QCD dynamics in the small-$x$ regime
and the  existence of very dense gluonic systems. In this regime, the probe can cover several nucleons inside the nucleus 
and interact with all of them {\it coherently}. Two approaches on the market treat this coherent kinematic
limit.  One of them is the Color Glass Condensate (CGC) 
approach~\cite{McLerran:1993ni,Albacete:2010bs,JalilianMarian:2011dt,Albacete:2012xq,Kang:2011ni},  which focuses on 
the non-linear corrections to QCD evolution equations.  It is only 
applicable at very small-$x$  and for transverse momenta  $\lesssim Q_s$, the saturation scale, where all the multiple 
scatterings are equally important. The other approach is the high-twist expansion approach, which treats the 
multiple scatterings as power suppressed corrections to the cross sections. It follows a generalized QCD 
factorization formalism~\cite{Qiu:1990xxa,Luo:1992fz,Luo:1993ui,Luo:1994np}  and 
computes the corrections order-by-order in a power series, where any additional 
correlated scattering is suppressed by an extra power of the momentum transfer. Within this approach,  in the 
forward rapidity limit, i.e. the proton direction, all nuclear size-enhanced power corrections~\cite{Qiu:2003vd,Qiu:2004da} 
to the differential  cross sections for both single hadron and dihadron production in p+A collisions have been resummed.   
When combined with cold nuclear matter energy loss~\cite{Neufeld:2010dz},  successful description of 
the single hadron suppression and dihadron correlation in the forward rapidity 
region has been demonstrated~\cite{Kang:2011bp}.

In this paper, we will focus on a different regime,  the region where the parton momentum fraction in the nucleus 
$x\sim {\cal O}(1)$ (outside small-$x$).
Incoherent  multiple interactions, relevant to the Cronin effect,  have been resummed before for uniform nuclear 
matter described by mean squared momentum transfer and parton scattering length~\cite{Gyulassy:2002yv}. 
Here, we follow the same generalized factorization theorem, discussed above, and attribute the first 
non-trivial multiple scattering  (double scattering) contributions to the twist-4 power-suppressed corrections to 
the  differential cross section.  
We demonstrate explicitly that in the $x\sim {\cal O}(1)$ regime only the incoherent multiple interactions are relevant. 
We take into account both initial-state and final-state double scattering effects and find that the final result 
can be written in terms of the well-defined twist-4 four-parton correlation functions. It depends on a universal
combination of second-derivative, 
first-derivative and non-derivative terms of these correlation functions that shares the same hard-scattering function.

The rest of our paper is organized as follows. In Sec.~II, we introduce our notation and study the double 
scattering contribution to the single inclusive hadron production in p+A collisions. Take one particular partonic 
subprocess $qq'\to qq'$ as an example to explain in detail how we derive the result and what are the  approximations 
we have used. In Sec.~III, we extend our method to the physical processes involving a photon. In particular, we 
study the double scattering contribution to the prompt photon production in p+A collisions, and 
to single hadron production in $\gamma$+A collisions. Both results  depend  on our findings from  Sec.~II. We 
summarize our paper in Sec.~IV.  We defer the  phenomenological  study, based on our result, to future publications.

\section{Multiple scattering effects for single inclusive hadron production  in p+A collisions}
\subsection{Single scattering contribution}
In this section we study single inclusive hadron production in p+A collisions, 
\bea
p(P')+A(P) \to h(P_h)+X,
\eea
where $P'$ is the momentum for the incoming proton, and $P$ is defined as the average momentum per nucleon 
in the nucleus. 
In general, the differential cross section for the above process can be expressed as a sum of contributions 
from single scattering, double scattering, and even higher multiple scattering \cite{Qiu:2001hj}:
\bea
d\sigma_{pA\to hX} = d\sigma_{pA\to hX}^{(S)} + d\sigma_{pA\to hX}^{(D)}  +\cdots,
\eea
where the superscript ``$(S)$'' indicates the single scattering contribution, and ``$(D)$'' represents the 
double scattering contribution. As illustrated in Fig.~\ref{generic}, in the single scattering contribution  
one parton $a$ from the proton interacts with one single parton~$b$ inside the nucleus to produce a parton $c$, 
which will then fragment into the final observed hadron $h$. On the other hand, in the double scattering contribution 
the same parton $a$ from the proton will interact with two partons $b, b'$ from the nucleus simultaneously 
to produce the final parton $c$. 
\bef
\psfig{file=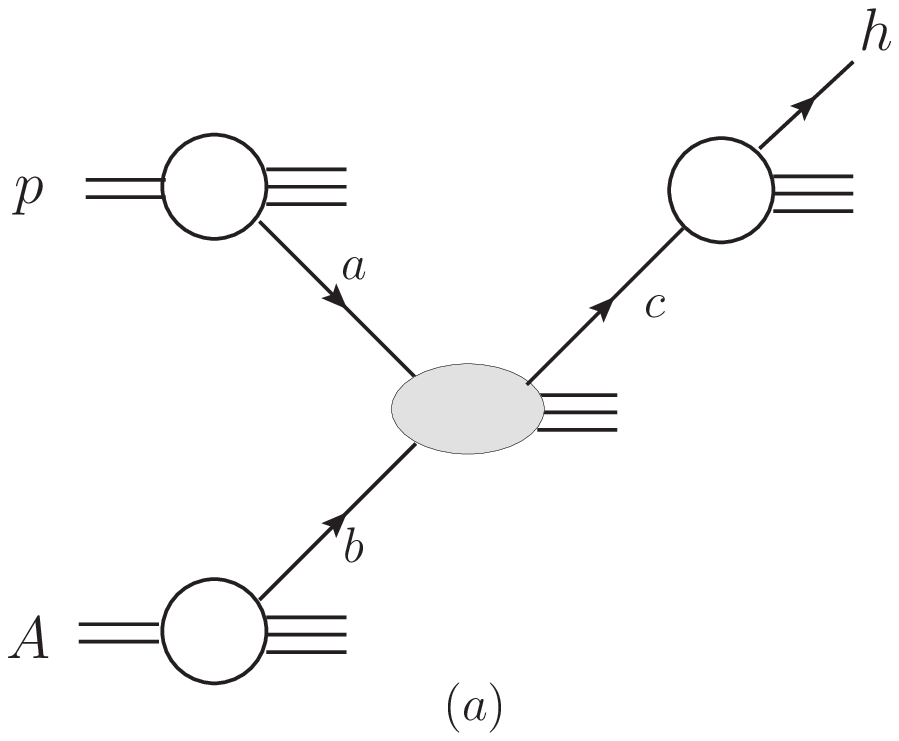, width=2.1in}
\hskip 0.3in
\psfig{file=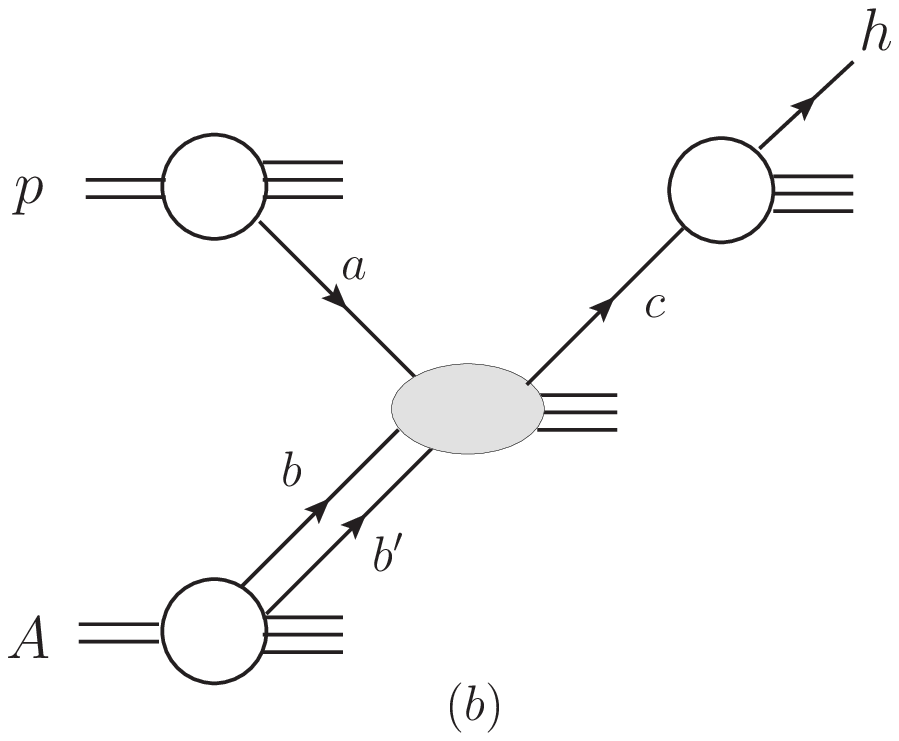, width=2.1in}
\caption{Generic diagrams for: (a) single scattering amplitude, where the parton $a$ from the proton interacts 
with one single parton $b$ inside the nucleus. (b) double scattering amplitude, where the parton $a$ from the proton 
interacts with two partons $b, b'$ from the nucleus simultaneously. Eventually, the hard scattering produces a 
parton $c$, which then fragments into the final observed hadron~$h$.}
\label{generic}
\eef

The single scattering contribution follows the usual leading-twist perturbative QCD factorization~\cite{Collins:1989gx}, 
and the differential cross section per nucleon at  leading order in the strong coupling $\alpha_s$ is given by
\bea
E_h\frac{d\sigma^{(S)}}{d^3P_h} = \frac{\alpha_s^2}{S} \sum_{a,b,c}\int \frac{dz}{z^2} D_{c\to h}(z) \int \frac{dx'}{x'} f_{a/p}(x')
\int \frac{dx}{x} f_{b/A}(x) H^U_{ab\to cd}(\hat s, \hat t, \hat u)\delta(\hat s+\hat t+\hat u),
\label{hadron}
\eea
where $\sum_{a,b,c}$ runs over all parton flavors, $S=(P'+P)^2$, $f_{a/p}(x')$ and $f_{b/A}(x)$ are the parton distribution 
functions inside the proton and the nucleus, respectively, and $D_{c\to h}(z)$ is the fragmentation function for parton $c$ 
transforming into hadron $h$. The scale dependencies are suppressed for brevity.  The hard-scattering functions 
$H^U_{ab\to cd}(\hat s, \hat t, \hat u)$ are the squared averaged matrix elements for the 
subprocess $a(p_a) + b(p_b) \to c(p_c) + d(p_d)$ with $p_a = x'P'$, $p_b = xP$, $p_c=P_h/z$ and the usual partonic Mandelstam variables:
\bea
\hat s = (x'P'+xP)^2,
\qquad
\hat t = (x'P'-p_c)^2,
\qquad
\hat u = (xP-p_c)^2.
\eea
These hard-scattering functions $H^U_{ab\to cd}(\hat s, \hat t, \hat u)$ are 
well-known~\cite{Owens:1986mp,Kang:2011bp,Kang:2010vd} and we reproduce them here for later convenience:
\bea
H^U_{qq'\to qq'}=&\frac{N_c^2-1}{2N_c^2}\left[\frac{{\hat s}^2+{\hat u}^2}{{\hat t}^2}\right],
\\
H^U_{qq\to qq}=&\frac{N_c^2-1}{2N_c^2}\left[\frac{{\hat s}^2+{\hat u}^2}{{\hat t}^2}
+\frac{{\hat s}^2+{\hat t}^2}{{\hat u}^2}\right]-\frac{N_c^2-1}{N_c^3}\left[\frac{{\hat s}^2}{{\hat t}{\hat u}}\right],
\\
H^U_{q\bar q\to q'\bar q'}=&\frac{N_c^2-1}{2N_c^2}\left[\frac{{\hat t}^2+{\hat u}^2}{{\hat s}^2}\right],
\\
H^U_{q\bar q\to q\bar q}=&\frac{N_c^2-1}{2N_c^2}\left[\frac{{\hat t}^2+{\hat u}^2}{{\hat s}^2}
+\frac{{\hat s}^2+{\hat u}^2}{{\hat t}^2}\right]-\frac{N_c^2-1}{N_c^3}\left[\frac{{\hat u}^2}{{\hat s}{\hat t}}\right],
\\
H^U_{qg\to qg}=& H^U_{gq\to gq} = -\frac{N_c^2-1}{2N_c^2}\left[\frac{{\hat s}}{{\hat u}}
+\frac{{\hat u}}{{\hat s}}\right]+\left[\frac{{\hat s}^2+{\hat u}^2}{{\hat t}^2}\right],
\\
H^U_{gq\to qg}=& H^U_{qg\to gq} = -\frac{N_c^2-1}{2N_c^2}\left[\frac{{\hat s}}{{\hat t}}
+\frac{{\hat t}}{{\hat s}}\right]+\left[\frac{{\hat s}^2+{\hat t}^2}{{\hat s}^2}\right],
\\
H^U_{q\bar q\to gg}=&\frac{(N_c^2-1)^2}{2N_c^3}\left[\frac{{\hat t}}{{\hat u}}
+\frac{{\hat u}}{{\hat t}}\right]-\frac{N_c^2-1}{N_c}\left[\frac{{\hat t}^2+{\hat u}^2}{{\hat s}^2}\right],
\\
H^U_{gg\to q\bar q}=&\frac{1}{2N_c}\left[\frac{{\hat t}}{{\hat u}}
+\frac{{\hat u}}{{\hat t}}\right]-\frac{N_c}{N_c^2-1}\left[\frac{{\hat t}^2+{\hat u}^2}{{\hat s}^2}\right],
\\
H^U_{gg\to gg}=&\frac{4N_c^2}{N_c^2-1}\left[3-\frac{{\hat t}{\hat u}}{{\hat s}^2}
-\frac{{\hat s}{\hat u}}{{\hat t}^2}-\frac{{\hat s}{\hat t}}{{\hat u}^2}\right],
\eea

\subsection{Double scattering contribution: $qq'\to qq'$ as an example}
Let us now concentrate on the double scattering contribution to the differential cross section. 
Since there are many partonic channels which contribute to the single inclusive hadron production, 
including $qq'\to qq'$, $qq\to qq$, $q\bar q\to q'\bar q'$, $q\bar q\to q\bar q$, $qg\to qg$, 
$gq\to gq$, $gq\to qg$, $qg\to gq$, $q\bar q\to gg$, $gg\to q\bar q$, and $gg\to gg$, we will take 
a simple partonic channel $qq'\to qq'$ as an example to demonstrate our method. The derivation for 
all other channels is similar. 

The double scattering diagrams can have both initial-state contributions, as shown in Fig.~\ref{initial}, and 
final-state contributions, as shown in Fig.~\ref{final}. The chosen process $qq'\to qq'$ is rather simple 
as only one Feynman diagram (the $t$-channel diagram which has one gluon exchange between the different quark flavors $q$ and $q'$) 
is relevant for the single scattering contribution. 
The Feynman diagrams shown in Figs.~\ref{initial} and \ref{final} are the complete set for double scattering contributions.
However, other partonic processes typically involve many more diagrams, which are all taken into account carefully 
in our calculation. Let us start with initial-state double scattering, 
for which there are three Feynman diagrams: in the first diagram Fig.~\ref{initial}(L), both gluons are on 
the left side of the cut line; in the second diagram Fig.~\ref{initial}(R), both gluons are on the right side 
of the cut line;  in the third diagram Fig.~\ref{initial}(M), the cut line is in the middle of the two gluons. 
Physically, Fig.~\ref{initial}(M) is the real diagram representing the classical double scattering picture, 
while both Figs.~\ref{initial}(L) and \ref{initial}(R) are the interference diagrams.
\bef
\psfig{file=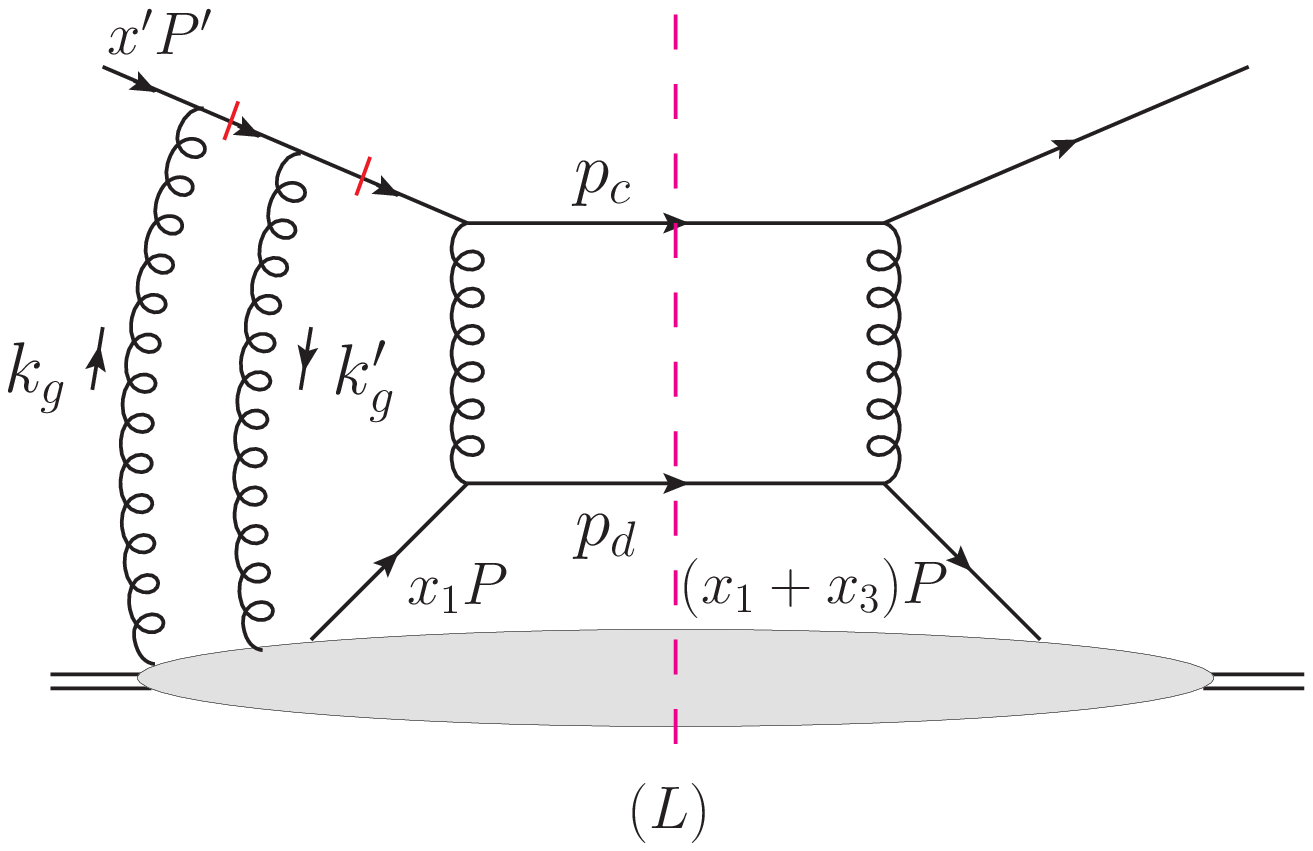, width=2.2in}
\hskip 0.1in
\psfig{file=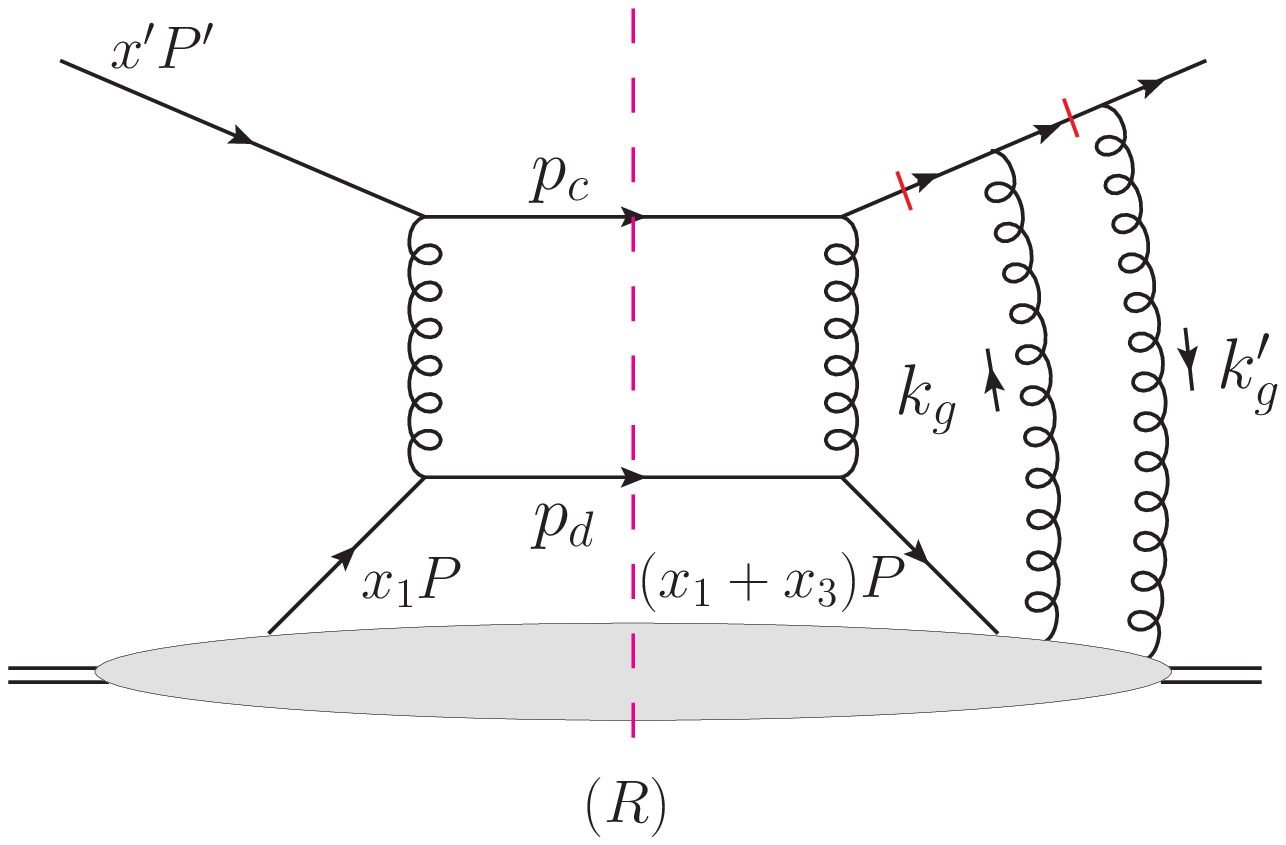, width=2.2in}
\hskip 0.1in
\psfig{file=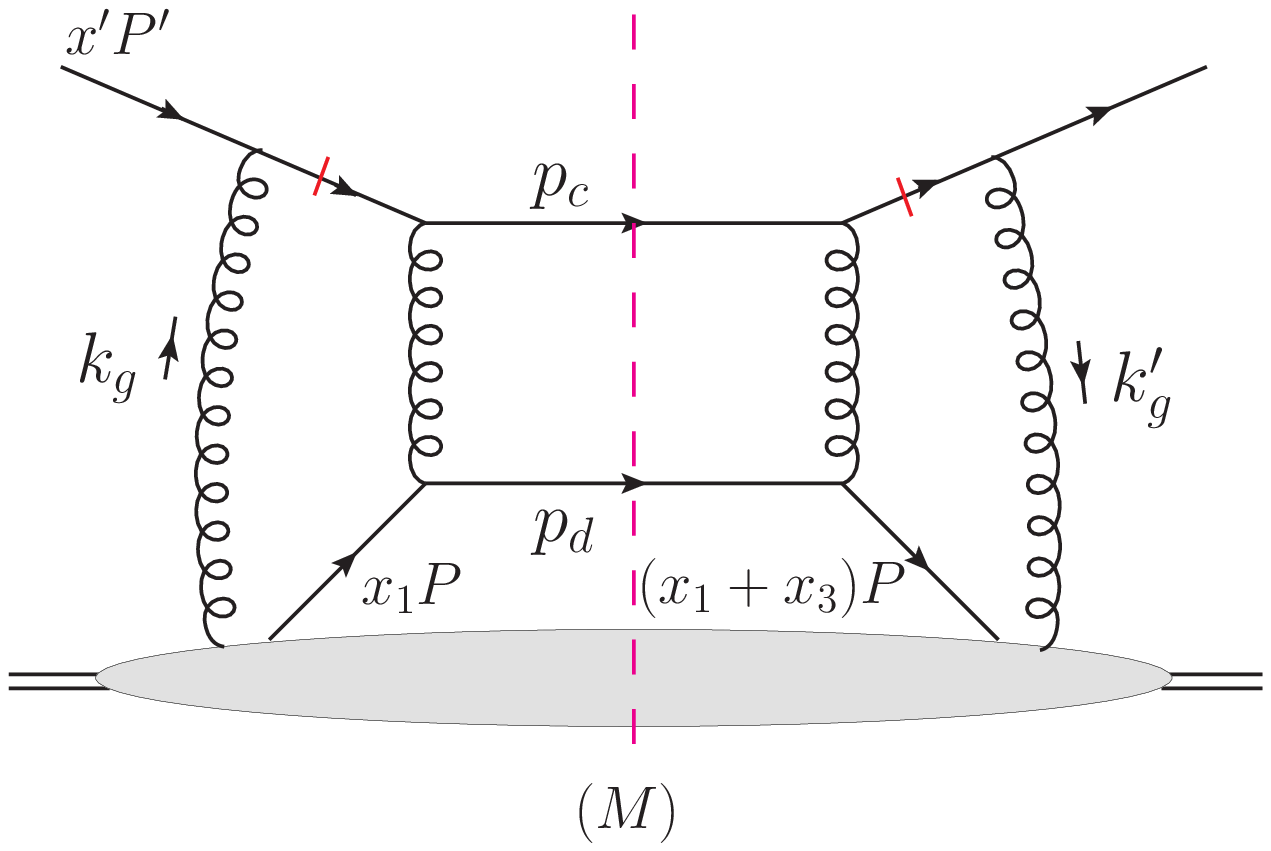, width=2.2in}
\caption{Initial-state double scattering contributions to the partonic subprocess $qq'\to qq'$. 
Here gluon momenta $k_g = x_2 P+k_\perp$ and $k_g' = (x_2 - x_3) P + k_\perp$.}
\label{initial}
\eef

These double scatterings  manifest themselves as twist-4 contributions to the differential cross section. 
To derive these contributions, a generalized factorization theorem, the so-called high-twist power expansion 
approach, was developed in~\cite{Qiu:1990xxa,Luo:1992fz,Luo:1993ui,Luo:1994np} some time ago. Since 
then, this approach has been used to study cold nuclear matter effects in either e+A or p+A collisions, 
such as energy loss, broadening effects, and dynamical nuclear shadowing. For examples beyond the ones
mentioned in the introduction, see~\cite{Guo:1998rd,Kang:2008us,Kang:2007nz,Kang:2012kc,Xing:2012ii}.

 Following this generalized factorization theorem, the double scattering contribution in Fig.~\ref{initial} 
can be expressed in terms of a twist-4 four-parton correlation function as follows:
\bea
E_h\frac{d\sigma^{(D)}}{d^3P_h} \propto \int \frac{dz}{z^2} D_{c\to h}(z) \int \frac{dx'}{x'} f_{a/p}(x')
\int dx_1 dx_2 dx_3 T(x_1, x_2, x_3) \left(-\frac{1}{2} g^{\rho\sigma}\right) 
\left[\frac{1}{2} \frac{\partial^2}{\partial k_\perp^\rho \partial k_\perp^\sigma} 
H(x_1, x_2,x_3, k_\perp)\right]_{k_\perp\to 0},
\label{master}
\eea
where $k_\perp$ is a small transverse momentum kick due to the multiple scattering 
and $T(x_1, x_2, x_3)$ is a twist-4 two-quark-two-gluon correlation function defined by
\bea
T(x_1, x_2, x_3) = \int \frac{dy^-}{2\pi} \frac{dy_1^-}{2\pi} 
\frac{dy_2^-}{2\pi} e^{ix_1P^+ y^-} e^{ix_2P^+(y_1^- - y_2^-)} e^{ix_3 P^+ y_2^-} \frac{1}{2} 
\langle P| F_{\alpha}^{~+}(y_2^-) \bar \psi_q(0) \gamma^+
\psi_q(y^-)F^{+\alpha}(y_1^-) |P\rangle.
\eea
$H(x_1, x_2,x_3, k_\perp)$ are the corresponding partonic hard-scattering functions, 
and the $x_1, x_2, x_3$ are the independent collinear momentum fractions carried by the partons 
from the nucleus, as shown in Fig.~\ref{initial}.

Here, the expansion around $k_\perp=0$ to  second order will extract the twist-4 contributions. 
This so-called collinear expansion is usually rather complicated and/or tedious in practice. In this paper, 
we will use a slightly improved technique for performing the collinear expansion and, thus, 
will be able to use it for single inclusive hadron production, which contains many partonic subprocesses. 
Such an improved technique was first developed for twist-3 expansion in studying transverse 
spin-dependent observables~\cite{Kouvaris:2006zy,Kang:2008ih}. It involves first integrating 
out the parton momentum fractions $x_1$, $x_2$, and $x_3$ with the help of either a kinematic $\delta$-function or 
contour integrals, then performing the $k_\perp$-expansion directly.  Though a small improvement, 
it enables us to perform the $k_\perp$-expansion with the help of the {\it Mathematica} package, 
instead of doing it by hand, as in the past.

We will now explain in detail how this works for our chosen example process, $qq'\to qq'$. 
We start with the first diagram Fig.~\ref{initial}(L). In this diagram, we have an on-shell 
condition for the unobserved quark $d$:
\bea
\delta(p_d^2) = & \delta\left[ \left(\left(x_1+x_3\right)P+x'P' - p_c\right)^2\right]
=  - \frac{x}{\hat t} \delta(x_1+x_3-x), 
\eea
which can be used to integrate out $x_1$ in Eq.~\eqref{master}, fixing $x_1 = x-x_3$. At the same time, 
there are propagators marked by a short bar in the diagram. These propagators are the so-called ``pole'' 
propagators, which will be used to perform contour integrals to eliminate the remaining two momentum fractions. 
They are given by the following expressions:
\bea
\frac{1}{(x'P'+x_2 P+k_\perp)^2+i\epsilon} =& \frac{x}{\hat s} \frac{1}{x_2 + x \frac{k_\perp^2}{\hat s} + i\epsilon},
\\
\frac{1}{(x'P'+x_3 P)^2 + i\epsilon} =& \frac{x}{\hat s} \frac{1}{x_3+i\epsilon}.
\eea
Now, the first propagator can be used to integrate out $x_2$,
\bea
\int dx_2 e^{i x_2 P^+(y_1^- - y_2^-)} \frac{1}{x_2 + x \frac{k_\perp^2}{\hat s} + i\epsilon} 
= -2\pi i\, \theta(y_2^- - y_1^-) e^{- i \frac{k_\perp^2}{\hat s} x P^+(y_1^- - y_2^-)},
\eea
which fixes $x_2 = -x \frac{k_\perp^2}{\hat s}$. On the other hand, the second propagator 
will be used to integrate out $x_3$,
\bea
\int dx_3 e^{ix_3 P^+ (y_2^- - y^-)} \frac{1}{x_3+i\epsilon} = -2\pi i\, \theta(y^- - y_2^-),
\eea
thus fixing $x_3 = 0$. Eventually, for both gluons on the left side of the cut line, we have the contribution proportional to
\bea
T_L(x_1, x_2, x_3) H_L(x_1, x_2, x_3, k_\perp),
\label{left}
\eea
with all the momentum fractions given by
\bea
x_1 = x, 
\qquad
x_2 = -x \frac{k_\perp^2}{\hat s}, 
\qquad
x_3 = 0,
\label{fractionLR}
\eea
and the relevant twist-4 correlation function 
\bea
T_L\left(x_1=x, x_2=-x \frac{k_\perp^2}{\hat s}, x_3=0\right) = 
& \int \frac{dy^-}{2\pi} \frac{dy_1^- dy_2^-}{2\pi} e^{ix P^+ y^-} e^{-i \frac{k_\perp^2}{\hat s} x P^+(y_1^- - y_2^-)}  
\theta(y_2^- - y_1^-) \theta(y^- - y_2^-)
\nnu
&\times
\frac{1}{2}
\langle P| F_{\alpha}^{~+}(y_2^-) \bar \psi_q(0) \gamma^+\psi_q(y^-)F^{+\alpha}(y_1^-) |P\rangle.
\label{leftT}
\eea

For the double scattering diagram  shown in Fig.~\ref{initial}(R), performing 
similar analysis, we have the contribution proportional to
\bea
T_R(x_1, x_2, x_3) H_R(x_1, x_2, x_3, k_\perp),
\label{right}
\eea
where the parton momentum fractions $x_1, x_2, x_3$ are the same as those in Fig.~\ref{initial}(L) and 
are given by Eq.~\eqref{fractionLR}. The relevant twist-4 correlation function is slightly different: 
\bea
T_R\left(x_1=x, x_2=-x \frac{k_\perp^2}{\hat s}, x_3=0\right) 
=& \int \frac{dy^-}{2\pi} \frac{dy_1^- dy_2^-}{2\pi} e^{ix P^+ y^-} e^{-i \frac{k_\perp^2}{\hat s} x P^+(y_1^- - y_2^-)}  
\theta(y_1^- - y_2^-) \theta(- y_1^-)
\nnu
&\times
\frac{1}{2}
\langle P| F_{\alpha}^{~+}(y_2^-) \bar \psi_q(0) \gamma^+\psi_q(y^-)F^{+\alpha}(y_1^-) |P\rangle.
\label{rightT}
\eea

Finally let us study the double scattering contribution in Fig.~\ref{initial}(M). 
In this case, the on-shell condition for the unobserved parton $d$ gives
\bea
\delta(p_d^2) =& \delta\left[\left((x_1+x_2)P+x'P'+k_\perp-p_c\right)^2\right]
= -\frac{x}{\hat t}\delta\left[x_1+x_2 - x - x\frac{k_\perp^2 - 2p_c\cdot k_\perp}{\hat t} \right],
\eea
which fixes $x_1 = x + x\frac{k_\perp^2 - 2p_c\cdot k_\perp}{\hat t} - x_2$. The two ``pole''
 propagators marked with short bars are given by the following expressions:
\bea
\frac{1}{(x'P'+x_2P+k_\perp)^2+i\epsilon} =& \frac{x}{\hat s} \frac{1}{x_2 + x \frac{k_\perp^2}{\hat s}+i\epsilon},
\\
\frac{1}{(x'P'+(x_2-x_3) P+k_\perp)^2-i\epsilon} =& - \frac{x}{\hat s} \frac{1}{x_3 - x_2 - x \frac{k_\perp^2}{\hat s}+i\epsilon},
\eea
which can be used to integrate over $x_2$ and $x_3$. Finally, we have
\bea
x_1 = x + x \frac{(k_\perp^2 - 2k_\perp\cdot p_c)}{\hat t}+ x \frac{k_\perp^2}{\hat s},
\qquad
x_2 = -x \frac{k_\perp^2}{\hat s},
\qquad
x_3 = 0.
\label{fractionM}
\eea
Thus for this diagram, the contribution can be written as
\bea
-T_M(x_1, x_2, x_3) H_M(x_1, x_2, x_3, k_\perp),
\label{middle}
\eea
with  momentum fractions $x_1, x_2, x_3$ given by Eq.~\eqref{fractionM}. The minus sign in front of 
the expression emphasizes the relative sign difference between Fig.~\ref{initial}(M) and 
Fig.~\ref{initial}(L,R) in the contour integration process.
The relevant correlation function $T_M(x_1, x_2, x_3)$ has different $\theta$-functions given by 
the following expression:
\bea
T_M\left(x_1=x + x \frac{(k_\perp^2 - 2k_\perp\cdot p_c)}{\hat t}+x \frac{k_\perp^2}{\hat s}, x_2 
= -x \frac{k_\perp^2}{\hat s}, x_3 = 0\right) =& \int \frac{dy^-}{2\pi} \frac{dy_1^- dy_2^-}{2\pi} 
e^{i \left(1 + \frac{(k_\perp^2 - 2k_\perp\cdot p_c)}{\hat t}+\frac{k_\perp^2}{\hat s} \right)x P^+ y^-} 
\nnu
&\times
e^{-i \frac{k_\perp^2}{\hat s} x P^+(y_1^- - y_2^-)}  
\theta(y^- - y_1^-) \theta(- y_2^-)
\nnu
&\times
\frac{1}{2}
\langle P| F_{\alpha}^{~+}(y_2^-) \bar \psi_q(0) \gamma^+\psi_q(y^-)F^{+\alpha}(y_1^-) |P\rangle.
\label{middleT}
\eea

The next critical step is to combine these three contributions, Eqs.~\eqref{left}, \eqref{right}, \eqref{middle}, 
and to perform the $k_\perp$-expansion:
\bea
\frac{\partial^2}{\partial k_\perp^\rho \partial k_\perp^\sigma} \Big[
T_L(x_1, x_2, x_3) H_L(x_1, x_2, x_3, k_\perp)+T_R(x_1, x_2, x_3) 
H_R(x_1, x_2, x_3, k_\perp)-T_M(x_1, x_2, x_3) H_M(x_1, x_2, x_3, k_\perp)\Big].
\label{expansion}
\eea
Here, we  use the following useful identity:
\bea
& \frac{\partial^2}{\partial k_\perp^\rho \partial k_\perp^\sigma} \big[T(x_1, x_2, x_3) H(x_1, x_2, x_3, k_\perp)\big]
 \nnu
&=
\frac{\partial^2 T}{\partial x_i \partial x_j} 
\left[\frac{\partial x_i}{\partial k_\perp^\rho} \frac{\partial x_j}{\partial k_\perp^\sigma} H \right] 
+ \frac{\partial T}{\partial x_i} 
\left[\frac{\partial^2 x_i}{\partial k_\perp^\rho \partial k_\perp^\sigma} H 
+ \frac{\partial x_i}{\partial k_\perp^\rho}\frac{\partial H}{\partial k_\perp^\sigma} 
+\frac{\partial x_i}{\partial k_\perp^\sigma} \frac{\partial H}{\partial k_\perp^\rho}\right]
+T \frac{\partial^2 H }{\partial k_\perp^\rho \partial k_\perp^\sigma},
\label{identity}
\eea
where repeated indices imply summation. With this identity in hand, we substitute the relevant parton 
momentum fractions $x_1, x_2, x_3$, given in Eqs.~\eqref{fractionLR} and \eqref{fractionM}, and then use 
{\it Mathematica} to perform the $k_\perp$-expansion automatically. For example, for the 
double-derivative term $\frac{\partial^2T(x_1, x_2, 0)}{\partial x_2^2}$, we obtain an expression 
proportional to 
\bea
\propto & \int \frac{dy^-}{2\pi} e^{i x P^+ y^-}  \int \frac{dy_1^- dy_2^-}{2\pi} (y_1^- - y_2^-)^2
\langle P| F_{\alpha}^{~+}(y_2^-) \bar \psi_q(0) \gamma^+\psi_q(y^-)F^{+\alpha}(y_1^-) |P\rangle
\nnu
&\times
H(x, 0, 0, 0)
\Big[\theta(y_2^- - y_1^-) \theta(y^- - y_2^-) + \theta(y_1^- - y_2^-) \theta(- y_1^-)
- \theta(y^- - y_1^-) \theta(- y_2^-)\Big],
\label{small}
\eea
where we have used the fact
\bea
H_L(x_1, x_2, x_3, k_\perp) = H_R(x_1, x_2, x_3, k_\perp) = H_M(x_1, x_2, x_3, k_\perp) \equiv H(x, 0, 0, 0)
\eea
for  $x_1 = x, x_2 = 0, x_3 = 0$, and $k_\perp = 0$. It is important to observe that in Eq.~\eqref{small} 
the factor
\bea
\Big[\theta(y_2^- - y_1^-) \theta(y^- - y_2^-) + \theta(y_1^- - y_2^-) \theta(- y_1^-)
- \theta(y^- - y_1^-) \theta(- y_2^-)\Big],
\label{contact}
\eea
is equivalent to the restrictions \cite{Luo:1992fz,Luo:1993ui,Luo:1994np}
\bea
|y^-| > |y_1^-| > |y_2^-|,
\label{constrain}
\eea
i.e., the integration $\int dy_1^- dy_2^-$ becomes an ordered integral limited by the value of $y^-$. 
In the region where the parton momentum fraction in the nucleus $x\sim {\cal O}(1)$, the rapidly oscillating exponential 
phase $e^{ixP^+y^-}$ will effectively restrict $y^-\sim 1/\left(xP^+\right) \to 0$, and 
thus also restricts $y_{1,2}^- \to 0$ 
through Eq.~\eqref{constrain}. Physically, this means that all the $y^-$ integrations in such a term are 
localized, and therefore, will not have nuclear size enhancement {\em for the double scattering contribution}. 
For this reason, such a term that 
is proportional to Eq.~\eqref{contact} is sometimes referred to as ``contact'' term \cite{Guo:2000nz}.

It is instructive and important to point out that in the small-$x$ region, where $x\to 0$, the above 
argument does not hold any more: in this case, even though one still has $y^-$ restricted to 
$y^-\sim 1/\left(xP^+\right)$, 
$y^-$ integration is not localized at $y^- \sim 0$ when $x\to 0$. Physically, this means that in the 
small-$x$ case,  the probe (incoming parton from the proton) can cover the whole nucleus. Thus, it will 
interact {\it coherently} with the partons from different nucleons at the same impact parameter inside 
the nucleus. In the small-$x$ region one cannot drop such contact term contributions as 
in Eq.~\eqref{contact}. In this regime  the  multiple scattering contributions to single hadron and 
dihadron production in p+A collisions were studied in the  forward rapidity limit 
in~\cite{Qiu:2003vd, Qiu:2004da}.  It has been shown that they lead to the so-called 
``dynamical shadowing'' effect, which has been 
used to successfully describe both single hadron suppression and dihadron correlation 
in the forward rapidity region  at the RHIC energies~\cite{Qiu:2004da,Kang:2011bp}.

In this paper, we will concentrate on the region where the parton momentum fraction in the nucleus $x\sim {\cal O}(1)$, i.e. 
outside the small-$x$ regime. 
We, thus, follow the original study \cite{Luo:1993ui,Luo:1994np} and neglect all these contact terms 
that are proportional to Eq.~\eqref{contact}. Finally, from Eqs.~\eqref{expansion} and~\eqref{identity}, 
we have the initial-state double scattering contribution to the $qq'\to qq'$ process as
\bea
\propto \left[x^2\frac{\partial^2 T^{(I)}_{q'/A}(x)}{\partial x^2} 
-x\frac{\partial T^{(I)}_{q'/A}(x)}{\partial x} +  T^{(I)}_{q'/A}(x)\right] c^{I}
H^{I}_{qq'\to qq'}(\hat s, \hat t, \hat u),
\label{qqini}
\eea
where the prefactor $c^I = -\frac{1}{\hat t} - \frac{1}{\hat s}$, the associated hard-scattering 
function $H^{I}_{qq'\to qq'} = C_F H^{U}_{qq'\to qq'}$, and the relevant two-quark-two-gluon correlation 
function $T^{(I)}_{q/A}(x)$ is given by \cite{Kang:2011bp,Kang:2008us,Xing:2012ii}
\bea
T_{q/A}^{(I)}(x) = &
 \int \frac{dy^{-}}{2\pi}\, e^{ixP^{+}y^{-}}
 \int \frac{dy_1^{-}dy_{2}^{-}}{2\pi} \,
      \theta(y^{-}-y_{1}^{-})\,\theta(-y_{2}^{-}) 
     \frac{1}{2}
     \langle P |F_{\alpha}^{\ +}(y_{2}^{-})\bar{\psi}_{q}(0)
                  \gamma^{+}\psi_{q}(y^{-})F^{+\alpha}(y_{1}^{-}) |P \rangle. 
\label{TqA}
\eea
The integration over $dy_1^-\,dy_2^-$ leads to the nuclear-size enhancement $\propto A^{1/3}$, 
as demonstrated in~\cite{Luo:1994np}. There are several comments in order. First, in the 
intermediate steps we have three independent two-quark-two-gluon correlation functions, which can 
be seen clearly from the different $\theta$-functions in Eqs.~\eqref{leftT}, \eqref{rightT}, and 
\eqref{middleT}. They are associated with the three different cuts, as shown in Fig.~\ref{initial}. 
However, only one correlation function $T_{q/A}^{(I)}(x)$ remains in the final result and it  
is associated with the diagram Fig.~\ref{initial}(M). Those associated with Figs.~\ref{initial}(L) 
and \ref{initial}(R) eventually lead to the contact combination, as in Eq.~\eqref{contact}. In other words, 
in the $x\sim {\cal O}(1)$ (outside the small-$x$) region, 
where  all the contact terms are suppressed,  the final result only depends on the real diagram shown 
in Fig.~\ref{initial}(M). The classical double scattering picture is preserved. 
It is because of this reason that the {\it coherent} nature in the multiple scattering disappears.
Thus, we should not expect to see the dynamical shadowing effect shown in \cite{Qiu:2003vd, Qiu:2004da}.

Second, the simple and compact form in Eq.~\eqref{qqini} is quite remarkable, i.e.
the final results for the combined second-derivative, first-derivative and non-derivative terms 
have a {\it common} hard-scattering function for this process, even though there could have  
been three separate hard-scattering functions multiplying 
$\frac{\partial^2 T^{(I)}_{q/A}(x)}{\partial x^2}$, $\frac{\partial T^{(I)}_{q/A}(x)}{\partial x}$, 
and $T^{(I)}_{q/A}(x)$, as in Eq.~\eqref{identity}. Similar simple structure was first 
observed in studying the transverse spin asymmetry at the twist-3 level~\cite{Kouvaris:2006zy,Kang:2008ih}, 
where one has only first-derivative and non-derivative terms and they share a single hard-scattering 
function. A more fundamental reason why this is the case deserves further investigation~\cite{Koike:2007rq}.

\bef
\psfig{file=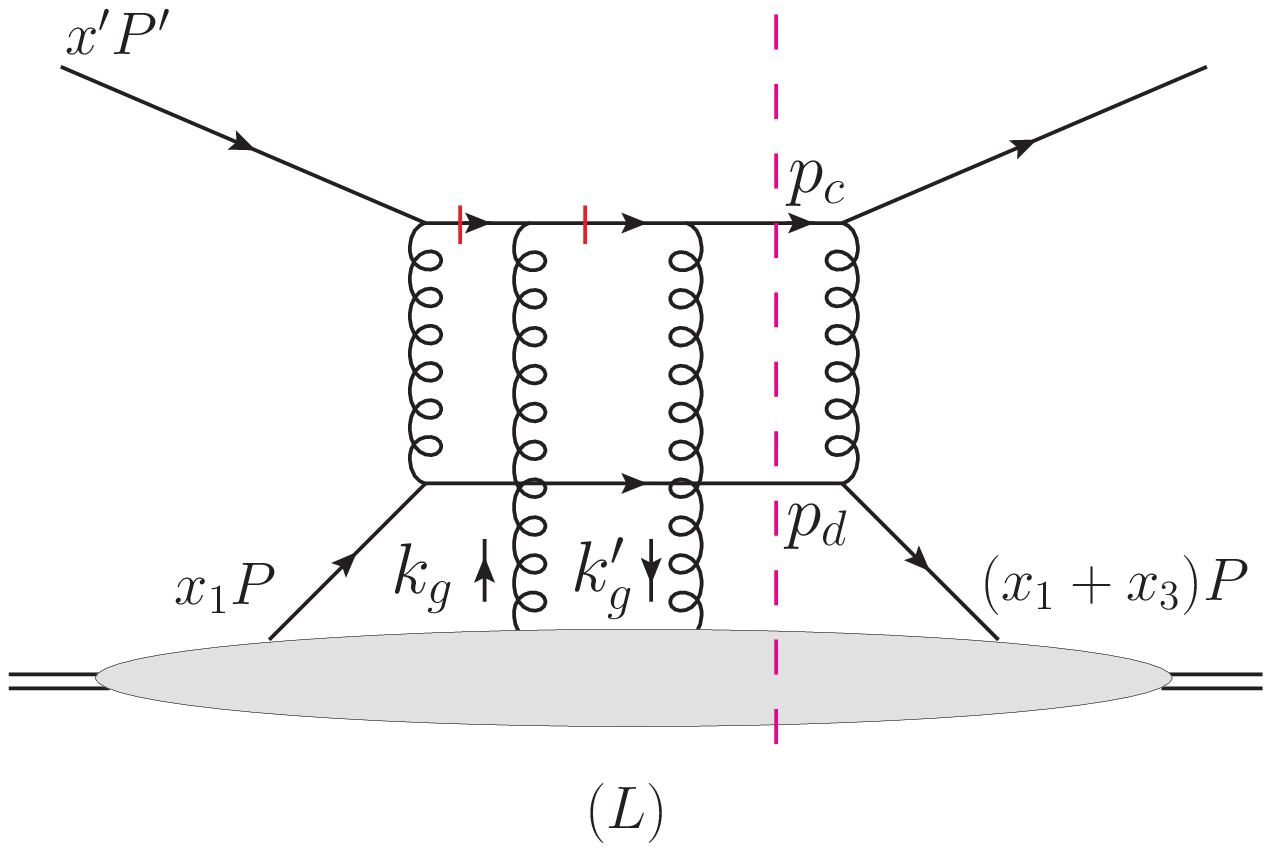, width=2.2in}
\hskip 0.1in
\psfig{file=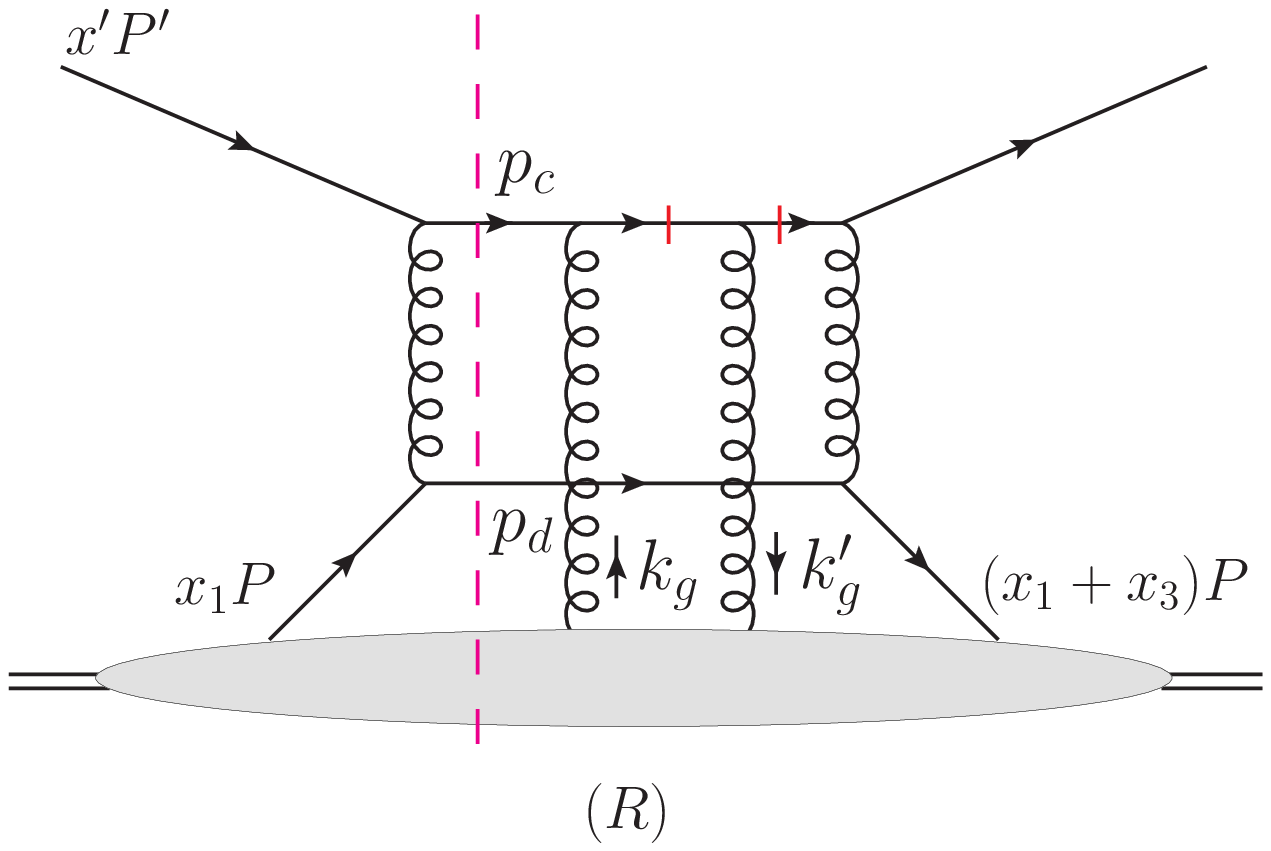, width=2.2in}
\hskip 0.1in
\psfig{file=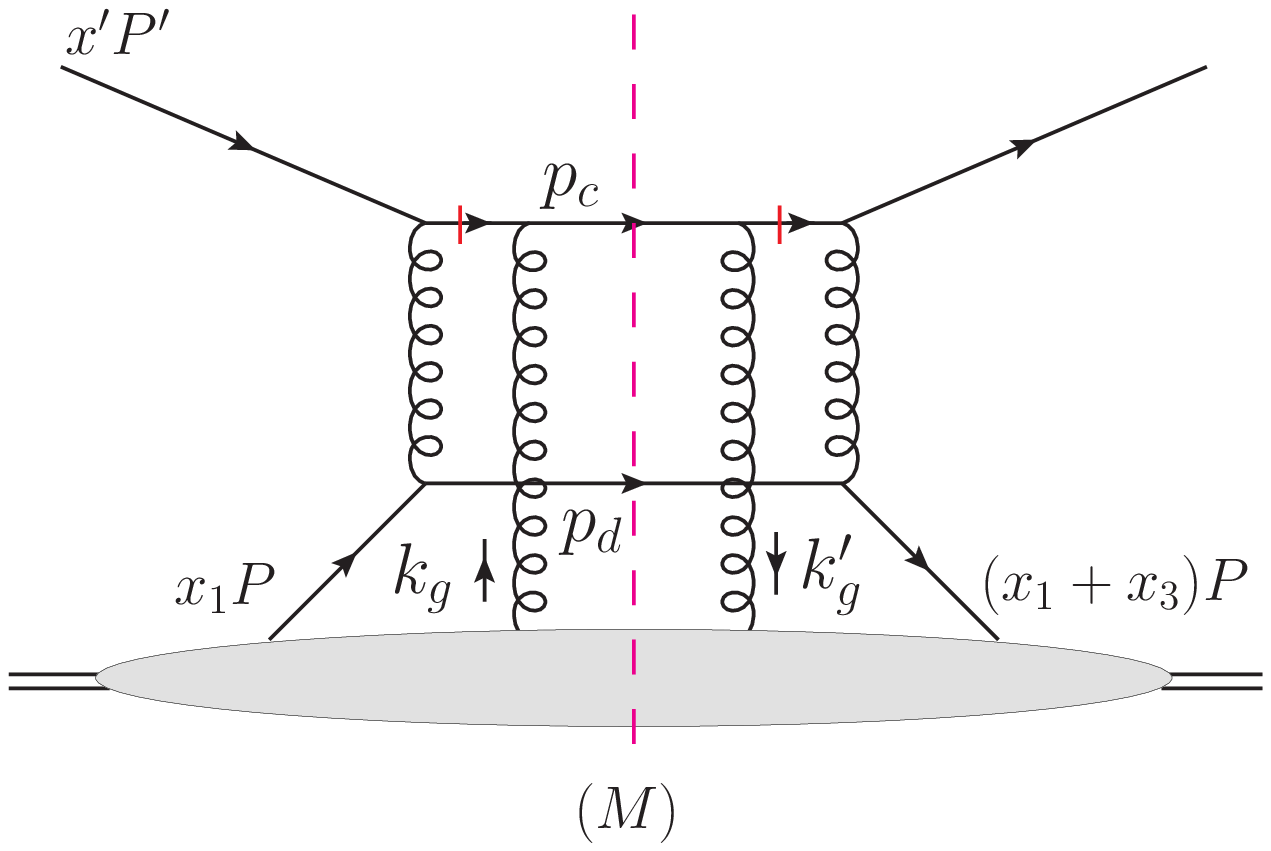, width=2.2in}
\caption{Final-state double scattering contributions to the partonic subprocess $qq'\to qq'$. 
Here, the gluon momenta $k_g = x_2 P+k_\perp$ and $k_g' = (x_2 - x_3) P + k_\perp$.}
\label{final}
\eef

Let us now turn to the final-state double scattering contributions to $qq'\to qq'$. 
The relevant Feynman diagrams are shown in Fig.~\ref{final}, where the {\it observed} 
outgoing parton $c$ undergoes double scattering (absorb soft gluons) directly. Following the 
same approach as above, the final result can again be written  in a compact form:
\bea
\propto \left[x^2\frac{\partial^2 T^{(F)}_{q'/A}(x)}{\partial x^2} 
-x\frac{\partial T^{(F)}_{q'/A}(x)}{\partial x} +  T^{(F)}_{q'/A}(x)\right] c^{F}
H^{F}_{qq'\to qq'}(\hat s, \hat t, \hat u),
\label{qqfin}
\eea
where we have a different prefactor $c^{F} =  -\frac{1}{\hat t} - \frac{1}{\hat u}$, and the 
final-state hard-scattering function is the same as the initial-state hard-scattering 
function as above $H^{F}_{qq'\to qq'} = C_F H^{U}_{qq'\to qq'}$. The relevant final-state 
two-quark-two-gluon correlation function $T^{(F)}_{q/A}(x)$ is the same as 
$T^{(I)}_{q/A}(x)$, except for the $\theta$-functions that are replaced as follows~\cite{Qiu:2003vd,Kang:2011bp,Kang:2008us,Xing:2012ii}
\bea
\theta(y^{-}-y_{1}^{-})\,\theta(-y_{2}^{-}) 
\to
\theta(y_{1}^{-}-y^{-})\,\theta(y_{2}^{-}).
\label{theta}
\eea
Here,  the final result again depends only on the real diagram Fig.~\ref{final}(M), 
 which preserves the classical double scattering picture. In principle, there are also 
Feynman diagrams in which the {\it unobserved} outgoing parton $d$ undergoes multiple scattering. 
The sum over different cuts for these diagrams will always lead to the contact term as in 
Eq.~\eqref{contact}, as demonstrated already in semi-inclusive deep inelastic 
scattering~\cite{Luo:1994np,Qiu:2003vd}. 
Thus, in the kinematic region $x\sim {\cal O}(1)$, we neglect such contributions.

\subsection{Final result: a compact form}
Likewise, we have computed both initial-state and final-state double scattering contributions 
to all other partonic channels: $qq\to qq$, $q\bar q\to q'\bar q'$, $q\bar q\to q\bar q$, 
$qg\to qg$, $gq\to gq$, $gq\to qg$, $qg\to gq$, $q\bar q\to gg$, $gg\to q\bar q$, and $gg\to gg$. 
For these processes, besides the two-quark-two-gluon correlation functions $T^{(I,F)}_{q/A}(x)$ 
defined above, four-gluon correlation functions $T^{(I,F)}_{g/A}(x)$ also contribute and they have 
the following definitions  \cite{Kang:2011bp,Kang:2008us,Xing:2012ii}:
\bea
T_{g/A}^{(I)}(x) = &
 \int \frac{dy^{-}}{2\pi}\, e^{ix P^{+}y^{-}}
 \int \frac{dy_1^{-}dy_{2}^{-}}{2\pi} \,
      \theta(y^{-}-y_{1}^{-})\,\theta(-y_{2}^{-}) 
\frac{1}{x P^+}\,
\langle P| F_\alpha^{~+}(y_2^-)
F^{\sigma+}(0)F^+_{~\sigma}(y^-)F^{+\alpha}(y_1^-)|P\rangle\, .
\label{TgA}
\eea
$T_{g/A}^{(F)}(x)$ is given by the same expression with the $\theta$-functions 
replacement specified in Eq.~\eqref{theta}. We find that the double scattering contributions for 
all these partonic processes follow the same compact form as in Eqs.~\eqref{qqini} and \eqref{qqfin}, 
with the following expression to the single hadron differential cross section:
\bea
E_h\frac{d\sigma^{(D)}}{d^3P_h} =&\left(\frac{8\pi^2\alpha_s}{N_c^2-1}\right) 
\frac{\alpha_s^2}{S} \sum_{a,b,c}\int \frac{dz}{z^2} D_{c\to h}(z) \int \frac{dx'}{x'} f_{a/p}(x')
\int \frac{dx}{x} 
\delta(\hat s+\hat t+\hat u)
\nnu
& \times \sum_{i=I, F}
\left[x^2\frac{\partial^2 T^{(i)}_{b/A}(x)}{\partial x^2} 
-x\frac{\partial T^{(i)}_{b/A}(x)}{\partial x} +  T^{(i)}_{b/A}(x)\right] c^{i}
H^{i}_{ab\to cd}(\hat s, \hat t, \hat u),
\label{main}
\eea
where $\sum_{i=I, F}$ denotes the summation over initial-state and final-state double scattering, $c^{i}$ are given by
\bea
c^{I} =& -\frac{1}{\hat t} - \frac{1}{\hat s},
\label{cI}
\\
c^{F} =&  -\frac{1}{\hat t} - \frac{1}{\hat u}.
\label{cF}
\eea
The hard-scattering functions $H^{i}_{ab\to cd}(\hat s, \hat t, \hat u)$ receive contributions
from both initial-state and final-state double scattering and are always proportional to the 
unpolarized hard-part functions $H^{U}_{ab\to cd}(\hat s, \hat t, \hat u)$ as follows:
\bea
H^I_{ab\to c d} =& \left\{
  \begin{array}{l l}
    C_F H^U_{ab\to c d} & \quad \text{a\ =\ quark}\\
     \\
    C_A H^U_{ab\to c d} & \quad \text{a\ =\ gluon}\\
  \end{array} \right.  \; , 
  \label{HI}
  \\
  \nnu
H^F_{ab\to c d} =& \left\{
  \begin{array}{l l}
    C_F H^U_{ab\to c d} & \quad \text{c\ =\ quark}\\
     \\
    C_A H^U_{ab\to c d} & \quad \text{c\ =\ gluon}\\
  \end{array} \right.   \; .
  \label{HF}
\eea
In other words, they only depend on the color of the incoming and outgoing partons 
which undergo the multiple scattering. It is important to emphasize again that the 
final result depends only on the diagrams in which the two gluons are on different 
sides of the  $t=\infty$ cut and preserve the classical double scattering picture. All 
the  interference diagrams drop out in the final result because they all show up in the 
contact terms and thus don't lead to the nuclear size enhancement in the 
$x\sim {\cal O}(1)$ (outside small-$x$) region, which we are interested in. Finally, if one 
replaces the hadron fragmentation function $D_{c\to h}(z)$ by $\delta(1-z)$, we immediately 
obtain the double scattering contribution to the single jet production in p+A collisions (to lowest order 
in the jet structure).

\section{Multiple scattering effects in physical processes involving a photon}
In this section we study the multiple scattering contributions to the physical processes 
which involve a photon in either the initial or the final state. In particular, we study 
the prompt  photon production in p+A collisions, and single inclusive hadron production 
in $\gamma$+A collisions. For both processes, our results derived in last section will 
be directly relevant, as we will show below.

\subsection{Multiple scattering in prompt photon production in p+A collisions}
The prompt photon production can receive two contributions: the so-called ``direct'' 
photons and ``fragmentation'' photons \cite{Gordon:1993qc,Vitev:2008vk,Gamberg:2012iq}. Thus, 
the  single scattering contribution to the prompt photon production can be written as:
\bea
E_\gamma \frac{d\sigma^{(S)}}{d^3P_\gamma} = \left.E_\gamma 
\frac{d\sigma^{(S)}}{d^3P_\gamma}\right|_{\rm direct} + \left. 
+ E_\gamma \frac{d\sigma^{(S)}}{d^3P_\gamma}\right|_{\rm frag}.
\eea
The direct photon contribution at the leading order has the following form:
\bea
\left.E_\gamma \frac{d\sigma^{(S)}}{d^3P_\gamma}\right|_{\rm direct} 
= \frac{\alpha_{em}\alpha_s}{S} \sum_{a,b} \int \frac{dx'}{x'} f_{a/p}(x')
\int \frac{dx}{x} f_{b/A}(x) H^U_{ab\to \gamma d}(\hat s, \hat t, \hat u)\delta(\hat s+\hat t+\hat u),
\eea
with the hard-scattering functions given by \cite{Owens:1986mp,Kang:2011rt}
\bea
H^U_{qg\to \gamma q}=&e_q^2 \frac{1}{N_c}\left[-\frac{{\hat s}}{\hat t}
-\frac{\hat t}{{\hat s}}\right],
\\
H^U_{gq\to \gamma q}=&e_q^2 \frac{1}{N_c}\left[-\frac{{\hat s}}{\hat u}
-\frac{\hat u}{{\hat s}}\right],
\\
H^U_{q\bar q\to \gamma g}=&e_q^2\frac{N_c^2-1}{N_c^2}\left[\frac{\hat t}{\hat u}
+\frac{\hat u}{\hat t}\right].
\eea
On the other hand, the single scattering contribution to the fragmentation photon production can be written as:
\bea
\left. E_\gamma \frac{d\sigma^{(S)}}{d^3P_\gamma}\right|_{\rm frag} 
= \frac{\alpha_s^2}{S} \sum_{a,b,c}\int \frac{dz}{z^2} D_{c\to \gamma}(z) \int \frac{dx'}{x'} f_{a/p}(x')
\int \frac{dx}{x} f_{b/A}(x) H^U_{ab\to cd}(\hat s, \hat t, \hat u)\delta(\hat s+\hat t+\hat u),
\label{fragphoton}
\eea
i.e., one replaces parton-to-hadron fragmentation function $D_{c\to h}(z)$ in 
Eq.~\eqref{hadron} by parton-to-photon fragmentation function $D_{c\to \gamma}(z)$. 

Let us now study the double scattering contributions to the prompt photon production. 
For the direct photon component, in which the photon is produced  in the hard scattering, 
we only have  initial-state double scattering. The calculation follows the same method 
as the last section and we have the result:
\bea
\left.E_\gamma \frac{d\sigma^{(D)}}{d^3P_\gamma}\right|_{\rm direct} =&\left(\frac{8\pi^2\alpha_s}{N_c^2-1}\right) 
\frac{\alpha_{em} \alpha_s}{S} \sum_{a,b} \int \frac{dx'}{x'} f_{a/p}(x')
\int \frac{dx}{x} 
\delta(\hat s+\hat t+\hat u)
\nnu
& \times
\left[x^2\frac{\partial^2 T^{(I)}_{b/A}(x)}{\partial x^2} 
-x\frac{\partial T^{(I)}_{b/A}(x)}{\partial x} +  T^{(I)}_{b/A}(x)\right] 
c^{I} H^{I}_{ab\to \gamma d}(\hat s, \hat t, \hat u),
\label{direct-photon}
\eea
where $c^I$ is given in Eq.~\eqref{cI}, and the associated hard-scattering functions $H^{I}_{ab\to \gamma d}$ are given by
\bea
H^I_{qg\to \gamma q} =& C_F H^U_{qg\to \gamma q},
\\
H^I_{gq\to \gamma q} =& C_A H^U_{gq\to \gamma q},
\\
H^I_{q\bar q\to \gamma g} =& C_F H^U_{q\bar q\to \gamma g}.
\eea
This initial-state double scattering contribution to direct photon production was first derived 
in~\cite{Guo:1995zk}.  Our approach allows for the  result to be written in a compact form, 
as in our Eq.~\eqref{direct-photon}.

In the fragmentation photon contribution Eq.~\eqref{fragphoton} 
one first produces a parton $c$ through the hard partonic process $ab\to cd$. This parton 
then fragments into the final observed photon. In this case, both the incoming parton $a$ and 
the outgoing parton $c$ can interact with the partons in the nucleus, resulting in both 
initial-state 
and final-state multiple scattering effects. 
These interactions are exactly the same as the ones in single inclusive hadron production, 
as shown in the previous section. We, thus, obtain the double scattering contribution to 
fragmentation photons as the following:
\bea
\left.E_\gamma\frac{d\sigma^{(D)}}{d^3P_\gamma}\right|_{\rm frag}
 =&\left(\frac{8\pi^2\alpha_s}{N_c^2-1}\right) \frac{\alpha_s^2}{S} \sum_{a,b,c}
\int \frac{dz}{z^2} D_{c\to \gamma}(z) \int \frac{dx'}{x'} f_{a/p}(x')
\int \frac{dx}{x}  \delta(\hat s+\hat t+\hat u)
\nnu
& \times \sum_{i=I, F}
\left[x^2\frac{\partial^2 T^{(i)}_{b/A}(x)}{\partial x^2} 
-x\frac{\partial T^{(i)}_{b/A}(x)}{\partial x} +  T^{(i)}_{b/A}(x)\right] c^{i}
H^{i}_{ab\to cd}(\hat s, \hat t, \hat u),
\eea
i.e., the final result is the same as Eq.~\eqref{main} with the replacement of the 
parton-to-hadron fragmentation function  with  the  parton-to-photon fragmentation function.

\subsection{Single inclusive hadron production in photon+nucleus collisions}
Let us now study the single hadron photo-production $\gamma+A\to h+X$, which can also receive 
two contributions: the so-called ``direct'' and ``resolved'' 
components~\cite{Owens:1979bn,Baer:1989jg,Frixione:1997ks,Harris:1997ji,Klasen:2002xb,Aurenche:1988vi}. 
The ``direct'' component corresponds to the case where the photon interacts directly 
with a parton in the nucleus. On the other hand, in the ``resolved'' contribution, 
the photon acts as a source of partons which collide with the partons in the nucleus. 
One can write the single scattering contribution to hadron photo-production as follows:
\bea
E_h\frac{d\sigma^{(S)}}{d^3P_h} = \left.E_h\frac{d\sigma^{(S)}}{d^3P_h}\right|_{\rm direct} 
+ \left.E_h\frac{d\sigma^{(S)}}{d^3P_h}\right|_{\rm resolved}.
\eea
The direct component can be written as
\bea
 \left.E_h\frac{d\sigma^{(S)}}{d^3P_h}\right|_{\rm direct}  
= \frac{\alpha_{em} \alpha_s}{S} \sum_{b,c}\int \frac{dz}{z^2} D_{c\to h}(z)
\int \frac{dx}{x} f_{b/A}(x) H^U_{\gamma b\to cd}(\hat s, \hat t, \hat u)\delta(\hat s+\hat t+\hat u),
\eea
where $S=(P_\gamma+P)^2$ is the center-of-mass energy squared, and the hard-scattering functions are given by \cite{Xing:2012ii}
\bea
H^U_{\gamma q\to qg} =& e_q^2 \frac{N_c^2-1}{N_c}\left[-\frac{\hat s}{\hat t} - \frac{\hat t}{\hat s}\right],
\\
H^U_{\gamma q\to gq} =& e_q^2 \frac{N_c^2-1}{N_c}\left[-\frac{\hat s}{\hat u} - \frac{\hat u}{\hat s}\right],
\\
H^U_{\gamma g\to q\bar q} =& e_q^2 \left[\frac{\hat t}{\hat u} + \frac{\hat u}{\hat t}\right].
\eea
On the other hand, the resolved component can be written as
\bea
\left.E_h\frac{d\sigma^{(S)}}{d^3P_h}\right|_{\rm resolved} = 
\frac{\alpha_s^2}{S} \sum_{a,b,c}\int \frac{dz}{z^2} D_{c\to h}(z) \int \frac{dx'}{x'} f_{a/\gamma}(x')
\int \frac{dx}{x} f_{b/A}(x) H^U_{ab\to cd}(\hat s, \hat t, \hat u)\delta(\hat s+\hat t+\hat u),
\eea
where $f_{a/\gamma}(x')$ is the parton distribution function in a photon, e.g. see Ref.~\cite{Baer:1989jg} for a parametrized functional form. 
All the partonic  subprocesses $ab\to cd$ are exactly the same as those in the single hadron 
production in p+A collisions,  with the same hard-scattering functions 
$H^U_{ab\to cd}(\hat s, \hat t, \hat u)$ as given in the last section.

Now, we turn our attention to the double scattering contribution to single hadron production 
in $\gamma$+A collisions, which can be studied experimentally in a future electron ion 
collider~\cite{Boer:2011fh}. For the direct component,  we have only a final-state double scattering 
contributions. The final result is
\bea
\left.E_h\frac{d\sigma^{(D)}}{d^3P_h}\right|_{\rm direct} 
=&\left(\frac{8\pi^2\alpha_s}{N_c^2-1}\right) \frac{\alpha_{em} \alpha_s}{S} 
\sum_{b,c}\int \frac{dz}{z^2} D_{c\to h}(z) 
\int \frac{dx}{x}  \delta(\hat s+\hat t+\hat u)
\nnu
& \times  \left[x^2\frac{\partial^2 T^{(F)}_{b/A}(x)}{\partial x^2} 
-x\frac{\partial T^{(F)}_{b/A}(x)}{\partial x} +  T^{(F)}_{b/A}(x)\right] c^{F}
H^{F}_{\gamma b\to cd}(\hat s, \hat t, \hat u),
\label{photo-production}
\eea
where $c^F$ is given in Eq.~\eqref{cF}, and the hard-scattering function $H^{F}_{\gamma b\to cd}$ 
is related to the unpolarized hard-part function $H^{U}_{\gamma b\to cd}$ just like in Eq.~\eqref{HF}:
\bea
H^F_{\gamma q\to qg} =& C_F H^U_{\gamma q\to qg},
\\
H^F_{\gamma q\to gq} =& C_A H^U_{\gamma q\to gq},
\\
H^F_{\gamma g\to q\bar q} =& C_F H^U_{\gamma g\to q\bar q}.
\eea

On the other hand, in the ``resolved'' component, the incoming particle is a parton resolved 
inside the photon  and it can interact with the nucleus via strong interaction. 
In this case, we have both initial-state and final-state multiple scattering effects. 
Again, the double scattering contributions will have the same form as in Eq.~\eqref{main}, except 
that $f_{a/p}(x')$ is replaced by $f_{a/\gamma}(x')$.  Finally,  if one replaces the fragmentation 
function $D_{c\to h}(z)$ by $\delta(1-z)$, one immediately obtains the double scattering contributions
to  single jet production in $\gamma$+A collisions at lowest order in the jet substructure. 
The double scattering contribution 
to the direct component for jet photo-production was first derived in~\cite{Luo:1992fz}. 
Our approach allows to write it in a simple compact form, Eq.~\eqref{photo-production}.

\section{Summary and discussion}

In this paper we studied the double scattering contribution to the differential cross section 
for single inclusive hadron production in p+A collisions within the high-twist factorization approach 
to parton interactions in cold nuclear matter. Unlike most recent studies,  
we concentrated on the region where the parton momentum fraction in the nucleus $x\sim{\cal O}(1)$. 
This regime,  outside small-$x$, represents the {\it incoherent} double scattering 
contribution that is most relevant at backward rapidities, i.e. in the direction of the nucleus. 
Including both initial-state and final-state double scattering contributions, 
we found that the final result is proportional  to a simple combination  of  second-derivative, 
first-derivative, and non-derivative terms of  
well-defined four-parton correlation functions that share the same hard-scattering functions. 
We further extended our method to study the double scattering contribution to prompt photon 
production in p+A collisions and to the single inclusive hadron production in $\gamma$+A 
collisions. All final results  follow  the same simple compact form, which is the main
finding of this work. 

We leave phenomenological studies for the future, since they require detailed modeling of the
four-parton correlation functions. Nevertheless, by direct inspection of our analytic results,  
we see that in the incoherent regime the double scattering  gives a positive 
contribution to the differential cross section for all processes considered here. Qualitatively, 
such  Cronin-like enhancement~\cite{Cronin:1974zm}  is directly comparable to the findings of  
alternative approaches to independent multiple parton scattering~\cite{Gyulassy:2002yv}. 
Owing to its power-suppressed nature, the nuclear effect is expected to disappear at large transverse 
momenta. At backward rapidities (i.e. in the direction of the nucleus) and transverse momenta up to a few GeV, 
p+A reaction always show enhancement of particle production relative to the naive binary collision-scaled
p+p result~\cite{Alber:1997sn,Abelev:2006pp,Adler:2004eh}. At midrapidity, the sign and magnitude 
of the nuclear enhancement depends on the center-of-mass energy. In the center-of-mass energy range up to 
5~TeV~\cite{ALICE:2012mj}, where measurements in p+A reactions exist, Cronin effect 
is present but its magnitude is significantly reduced in going from the fixed target experiments to 
RHIC and, finally, to the LHC. We expect that our work will shed light on the origin of cold nuclear 
matter effects in the unexplored backward rapidity region in both p+A and e+A reactions. It will also 
help understand the transition from incoherent to coherent multiple scattering 
effects~\cite{Vitev:2006bi} at forward rapidity.

\section*{Acknowledgments}
Z.K. thanks J.-W. Qiu for helpful discussions. This work is supported by the US 
Department of  Energy, Office of Science and in part by the LDRD 
program at LANL and NSFC of China under Project No. 10825523.


\end{document}